\documentclass[preprint,12pt,floatfix,showpacs,showkeys]{revtex4-1}
\usepackage{amsmath,bm,graphicx}
\usepackage{amsmath}
\usepackage{amssymb}
\usepackage{amsfonts}
\usepackage{graphicx,color}

\begin{document}
\title{Hydrodynamic signatures of stationary Marangoni-driven surfactant transport}
\author{M. M. Bandi}
\email[Co-corresponding author:]{ bandi@oist.jp}
\affiliation{Collective Interactions Unit, OIST Graduate University, Okinawa, Japan, 904-0495}

\author{V. S. Akella}

\author{D. K. Singh}
\affiliation{Collective Interactions Unit, OIST Graduate University, Okinawa, Japan, 904-0495}

\author{R. S. Singh}
\affiliation{School of Engineering, Brown University, Providence, Rhode Island 02912}

\author{S. Mandre}
\email[Co-corresponding author:]{ shreyas\_mandre@brown.edu}
\affiliation{School of Engineering, Brown University, Providence, Rhode Island 02912}

\date{\today}

\begin{abstract}
We experimentally study steady Marangoni-driven surfactant transport on the interface of a deep water layer. 
Using hydrodynamic measurements, and without using any knowledge of the surfactant physico-chemical properties, we show that sodium dodecyl sulphate and Tergitol 15-S-9 introduced in low concentrations result in a flow driven by adsorbed surfactant.
At higher surfactant concentration, the flow is dominated by the dissolved surfactant.
Using Camphoric acid, whose properties are {\it a priori} unknown, we demonstrate this method's efficacy by showing its spreading is adsorption dominated. 
\end{abstract}

\maketitle

Surfactants introduced at liquid interfaces give rise to Marangoni stresses that drive a flow \cite{Levich1969}. 
The fundamental process of surfactant spreading, governed by its diffusion and transport {\it via} self-induced flow has many applications from materials chemistry to biomechanics \cite{Jobe1993,Grotberg1994,Harshey2003,Eisner2005,Bush2006,Cantat2013,Ivanov1997,Barnes2008,Herzig2011}. 
Many surfactants are soluble in the fluid and could be transported in a phase dissolved in the bulk or adsorbed at the interface \cite{Roche2014,LeRoux2016}. 
A complete description of the resulting flow is hindered by the complexity of surfactant dynamics, which includes characterizing the equilibrium adsorption characteristics, the adsorption-desorption kinetics, and the transport by the flow \cite{Chang1995,Eastoe2000}.
Whereas methods based on molecular \cite{Lu2000} or radiometric \cite{Salley1950,Matuura1958,Tajima1970} markers can measure surface excess during a flow \cite{Manning1998}, low surfactant diffusivity into bulk fluid renders bulk concentration measurements at the interface difficult during flow. 
Direct Marangoni stress measurements {\it via in situ} surface tension gradient measurements are equally challenging. 
Simultaneous access to bulk and surface concentrations, Marangoni stress, sorption kinetics, and their subsequent correlation with one another to deduce the surfactant dynamics remains a formidable task.

In a recent study \cite{Roche2014}, for example, surfactant was introduced on the air-water interface through a steady point source. 
Simple scaling laws for surfactant spreading were derived by assuming the sorption kinetics to be much faster than the hydrodynamics so that the dynamics were dominated by the dissolved phase. 
Verification of this assumption was not possible owing to the aforementioned difficulties.
A possible alternative is that the sorption kinetics are too slow compared to the hydrodynamics, so that the dynamics are governed by the adsorbed phase.
Either of these assumptions reduce the complexity of the problem by enabling semi-analytical steady solutions to the governing equations \cite{Mandre2017b}.
Our objective in this setting is an experimental validation of these assumptions using hydrodynamic measurements alone.

\begin{figure}
\begin{center}
\includegraphics[width=0.48\textwidth]{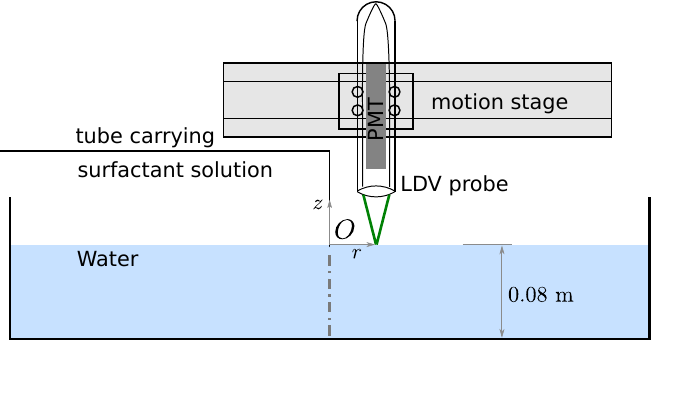}
\end{center}
\caption{(Color online) Schematic of the experimental setup.}
\label{fig:Schematic}
\end{figure}

Consider a surfactant released steadily on the interface through a source much smaller in radial extent than the container size (see Figure \ref{fig:Schematic}) such that a steady axisymmetric flow is established (see Supplemental Material - Movie M1 for visualization). 
In the region much larger than the source but much smaller than the container, the source may be idealized as a point and the container assumed infinite. 
Furthermore, consider the fluid viscosity and surfactant diffusivity to be small enough that most of the flow and surfactant concentration is established within a boundary layer near the surface.
These approximations, along with the assumption of adsorption- or dissolution-dominated surfactant dynamics, render the governing physical description scale invariant. 
Consequently, the fluid radial $u(r,z)$ velocity components in cylindrical coordinates $(r,z)$, exhibit a self-similar structure \cite{Mandre2017b}. 
Three experimentally measured invariant characteristics of this self-similar flow serve as hydrodynamic signatures of the simplified surfactant transport.

In this letter, we present experimental verification of these flow signatures using two generic surfactants in water -- sodium dodecyl sulphate (SDS) and Tergitol 15-S-9 (Tergitol).
Both these surfactants are water soluble (solubility 0.2 kg/l and 0.7 kg/l, respectively) and span a range of critical micellar concentration (CMC) from $8 \times 10^{-5}$ mM for Tergitol to $8 \times 10^{-3}$ mM for SDS.
SDS is ionic in nature, while Tergitol is non-ionic.
Without using any knowledge of the surfactant physico-chemical parameters, we show that for concentrations less than 15\% CMC, both surfactants exhibit flows dominated by adsorbed surfactant. 
In the same manner, mixture concentrations between $24-50\%$ CMC exhibit flow dominated by the dissolved surfactant.
Finally, we also determine which of the two processes dominate the dynamics of a third surfactant, camphoric acid (CA), released at the interface from a gel tablet at unknown rates and concentrations.

\begin{figure}
\begin{center}
\includegraphics[width=0.48\textwidth]{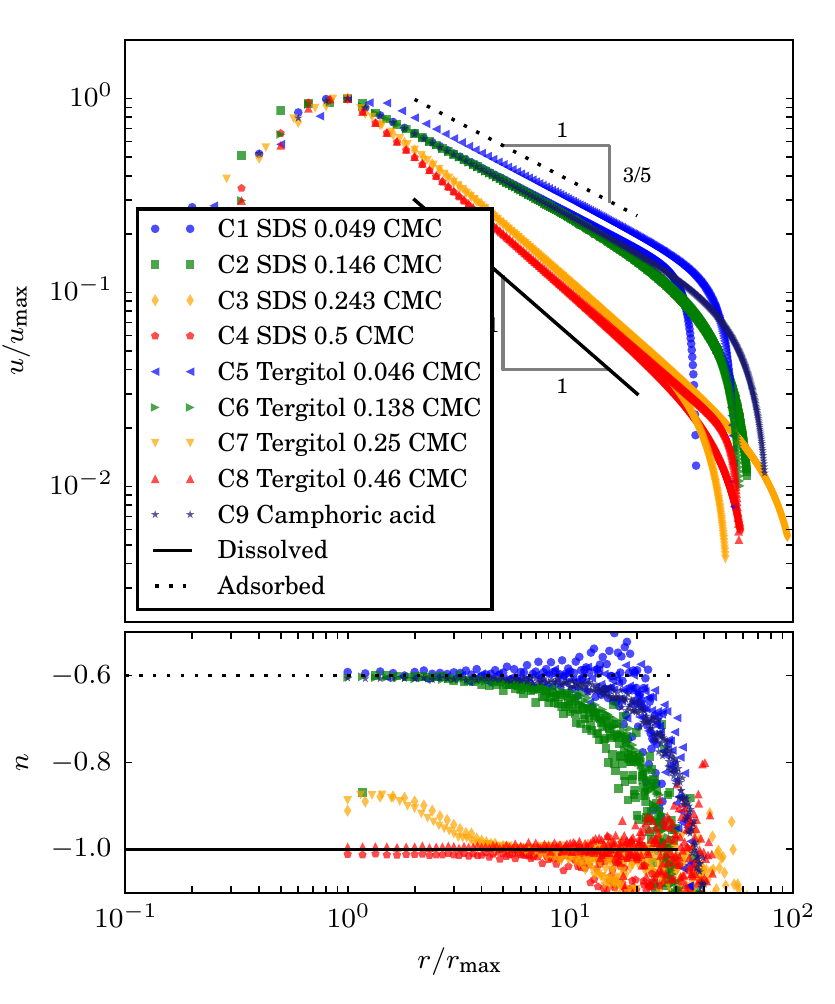}
\end{center}
\caption{(Color online) (a) Radial velocity component $u(r,0)$ at the fluid surface as a function of distance from the source center for three surfactants.
Also plotted are power laws \eqref{eqn:adspowerlaw} and \eqref{eqn:dispowerlaw} expected for the dissolution (solid black line) and adsorption (dashed black line) dominated cases. 
The velocity is rescaled by its maximum value $u_\text{max}$ on the interface, and $r$ is rescaled by $r_\text{max}$, the location where the maximum velocity occurs. 
(b) Same data as (a), but presented in the form of power law exponent $n = d (\log u)/d (\log r)$.}
\label{fig:expplots}
\end{figure}

Solution of SDS or Tergitol was introduced on air-water interface {\it via} a borosilicate capillary (tip inner diameter of 3-5 $\mu$m) by Marangoni suction, a procedure empirically determined to minimize forcing a radial jet due to hydrodynamic pumping \cite{Squire1951,Landau1959,Laohakunakorn2013}. 
Four different concentrations for SDS and Tergitol ranging from about $0.05$ CMC to $0.5$ CMC (labeled C1-8 in Figure \ref{fig:expplots}) were used to span the range of surfactant dynamics from adsorption-dominated to dissolution-dominated.
CA was introduced on the interface through an agarose gel tablet (diameter 3 mm, thickness 1 mm) infused with CA (case C9 in Figure \ref{fig:expplots}). 
The gel tablet was mounted on a vertical motion stage and brought in contact with the interface. 
In our experiments, the velocity boundary layer was minimally influenced by the dish bottom. 
The velocity profiles $u(r,0)$ and $u(r=r_1, z)$, and the surface shear $u_z(r, z=0)$ of the axisymmetric flow that developed due to the Marangoni flow were measured using Laser Doppler Velocimetry (LDV). 

The reproducibility required for the experiment and the measurement precision in velocity up to 4$^\text{th}$ decimal place to ascertain the power laws and the boundary layer profile reported here require a tight protocol (for full experimental details, please see Supplemental Material).

{\it Signature 1:} The measured surface radial velocity $u(r,0)$ is shown on a logarithmic scale in Fig. \ref{fig:expplots}(a). 
A correction to account for higher order effects due to finite size of the CA tablet is applied, as detailed in Supplemental Materials. 
For all the nine cases considered,  $u(r,0)$ reaches a maximum $u_\text{max}$ at $r = r_\text{max}$ (about 1 mm), and decays approximately as a power law in a range of radii $1 < r/r_\text{max} < 20$.
For $r/r_\text{max} \gtrsim 20$, $u(r,0)$ decreases much faster than the power-law decay. 

The exponent of the power-law decay in the intermediate range of radii is the first hydrodynamic signature of the surfactant dynamics.
In this range, five of the nine cases (C1, C2, C5, C6 and C9), those with surfactant concentrations $<0.15$ CMC and the one with CA, exhibit an approximate decay  of $u(r,0)$ as $r^{-3/5}$.
(For enhanced visibility, cool colors depict these cases in Figures \ref{fig:expplots} and \ref{fig:signature3}.)
The remaining four cases (C3, C4, C7 and C8), which include surfactant concentration $>0.24$ CMC (shown in warm colors), exhibited decay as $r^{-1}$.

To confirm the measured slopes, Fig. \ref{fig:expplots}(b) plots the log derivative (Selke's method \cite{Selke1991}) $n=\text{d}~\log u/ \text{d}~\log r$ as a function of $r$. 
The differentiation is performed using finite differences between neighboring experimentally measured data points. 
For the lowest concentrations of the SDS (0.049 CMC) and Tergitol (0.046 CMC), and in the range $1<r/r_\text{max}<10$, the value of $n$ lies between -0.565 and -0.618.
For the next lowest concentration (0.146 CMC for SDS And 0.138 CMC for Tergitol), $n$ departs from this range at $r/r_\text{max} \gtrsim 8$. 
As the concentration is increased further (0.243 CMC for SDS and 0.25 CMC for Tergitol), $n$ lies in the range $-0.87$ to $-1.04$, with a systematic departure from $\approx -1$ occuring in the range $1<r/r_\text{max} \lesssim 4$.
And finally, for the largest concentration (0.5 CMC for SDS and 0.46 CMC for Tergitol), $n$ lies in the range $-0.98$ to $-1.03$.
For the flow driven by CA, $n$ lies in the range $-0.60$ to $-0.63$.
Based on these oservations, we posit two values for the power-law exponents, $n\approx -0.6$ and $n\approx -1$, with the random variation attributed to measurement noise and the systematic deviations to departures from the asymptotic regimes of validity.
\begin{figure}
\begin{center}
\includegraphics[width=0.48\textwidth]{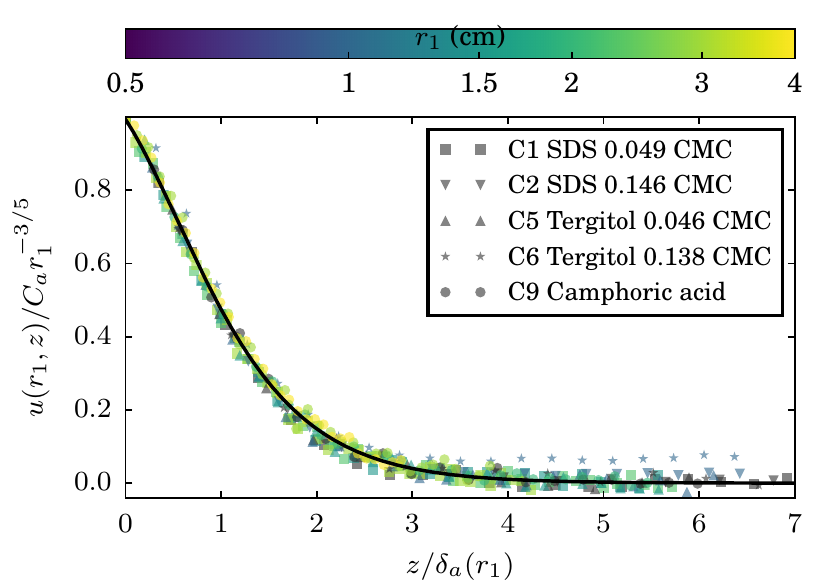}
\end{center}
\caption{The radial velocity profile in the boundary layer for flow dominated by adsorbed surfactant. The experimentally measured radial velocity profile (symbols) is normalized according to \eqref{eqn:uansatz} and plotted against the similarity coordinate. Also plotted (solid curve) is the self-similar profile derived theoretically by solving \eqref{eqn:selfsimilarode}.}
\label{fig:cabl}
\end{figure}

These power laws can be understood in terms of the competing fluid and surfactant-induced stresses as follows. 
Due to self-similar nature of the flow, the length scale in the radial and depth-wise directions are $r$ and the boundary layer thickness, $\delta(r)$, respectively. 
Fluid inertia scales as $\rho u^2/r$ ($\rho$ is fluid density) while viscous forces scale as $\mu u /\delta^2$ ($\mu$ is dynamic viscosity). 
A balance between the two is expected in the boundary layer, which furnishes one relation, $\delta \sim \sqrt{\mu r/\rho u}$. 
Imposing the Marangoni stress, which scales as $\Delta \sigma/r$ ($\Delta \sigma$ being the reduction in surface tension) to be equal to the scale of the fluid's shear stress, $\mu u/\delta$, leads to $\delta = \mu u r/\Delta \sigma$. 
The two cases are distinguished by the relation between $\Delta \sigma$ and the surfactant concentration, and how the surfactant is transported.

When the surfactant dynamics are dominated by the adsorbed phase, surfactant conservation implies $2 \pi u r c_2 = q_2$, where $c_2$ is the surface concentration of the surfactant and $q_2$ its surface flux. 
Here we neglect the diffusion of surfactant. The surface tension depends on surfactant concentration as $\Delta \sigma = -\Gamma_2 c_2$, where $\Gamma_2$ is a proportionality constant. 
Eliminating $c_2$ and $\Delta \sigma$ leads to 
\begin{align}
u(r,z=0) =f'(0) C_a r^{-3/5},\quad \delta_a(r) = r^{4/5} \sqrt{\nu/C_a},
\label{eqn:adspowerlaw}
\end{align}
where $C_a=(\Gamma_2^2 q_2^2 \nu/ (4\pi^2\mu^2))^{1/5}$, $f'(0)$ is a dimensionless proportionality constant to be determined, and $\nu=\mu/\rho$. 
\begin{figure}
\begin{center}
\includegraphics[width=0.48\textwidth]{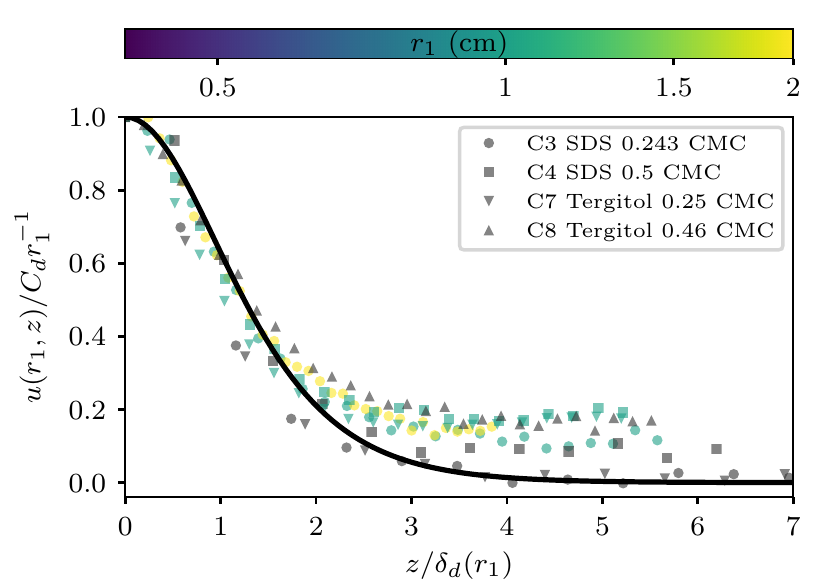}
\end{center}
\caption{The radial velocity profile in the boundary layer dominated by dissolved surfactant. The experimentally measured radial velocity profile (symbols), normalized by the scaling according to \eqref{eqn:u2ansatz}, plotted against the similarity coordinate. Also plotted (solid curve) is the self-similar profile from \eqref{eqn:u2ansatz}.}
\label{fig:cabl2}
\end{figure}

When surfactant dynamics are dominated by the dissolved phase, surfactant bulk concentration $c_3(r,z)$ obeys an advection-diffusion equation with diffusivity $D$, and $\Delta \sigma = -\Gamma_3 c_3$, where $\Gamma_3$ is a material-dependent constant. 
The surfactant diffuses in a boundary layer of thickness $\delta_c = \sqrt{Dr/u}$, and hence surfactant conservation implies $2 \pi r u c_3 \delta_c \propto q_3$, where $q_3$ is the volumetric surfactant release rate, which yields $c_3 \propto q_3/\sqrt{u r^3 D}$. 
The resulting Marangoni stress scales as $\Gamma_3 c_3/r \propto \Gamma_3 q_3 /\sqrt{u r^5 D}$, which balances the fluid viscous shear stress. 
The shear stress at the surface, due to a peculiarity in the boundary layer flow structure, does not scale as $\mu u/\delta$, but scales one order weaker in the small parameter $\delta/r$, as $\mu u/r$. 
Balancing the scales for Marangoni stress and shear stress yields
\begin{align}
u(r,z=0) = C_d r^{-1}, \quad \delta_d(r) = r \sqrt{\nu/C_d}
\label{eqn:dispowerlaw}
\end{align}
where $C_d = {(\Gamma_3^2 q_3^2/(8\pi^3 \mu^2 D))^{1/3}}$. 
These scaling estimates and the appropriate dimensionless proportionality constant are determined from an exact similarity solution by Bratukhin and Maurin \cite{Bratukhin1967} \cite[for details see][]{Mandre2017b}.

{\it Signature 2:}
To ensure that the power law exponents arise due to the fluid dynamics presented here, and not due to any unexpected coincidences, we compare the depth-wise profile $u(r_1,z)$ with theoretical expectations.
In the case of adsorption-dominated surfactant dynamics, the solution may be expressed as 
\begin{align}
u(r,z) &= C_a r^{-3/5} f'(\xi), \label{eqn:uansatz} 
\end{align}
in terms of a similarity coordinate $\xi = z/ \delta_a(r)$ and a self-similar profile $f(\xi)$.
Here $f$ satisfies \cite{Mandre2017b}
\begin{align}
 f'''(\xi) + \dfrac{3}{5} f'(\xi)^2  + \dfrac{6}{5} f(\xi)f''(\xi) = 0, \label{eqn:selfsimilarode}
\end{align}
and $f(0) = 0$,  $f''(0) f'(0) = \dfrac{2}{5}$, and  $f'(-\infty) = 0$.
This third order ordinary differential equation is solved using a shooting method to obtain $f$ and the $u(r,z)$ is re-constructed using \eqref{eqn:uansatz}.
The proportionality constant, $f'(0)\approx  0.9943$ in \eqref{eqn:adspowerlaw}, is obtained as part of this solution.

Similarly, a leading order approximation to the boundary layer flow profile driven by the surfactant whose dynamics are dominated by the dissolved phase \cite{Mandre2017b} is
\begin{align}
u(r, z) = C_d r^{-1} \text{sech}^2 \left( \dfrac{z}{\delta_d(r) \sqrt{2} } \right) + O\left(\dfrac{\delta_d(r)}{r}\right).
\label{eqn:u2ansatz} 
\end{align}

Figure \ref{fig:cabl} and \ref{fig:cabl2} show a comparison of experimentally measured depth-wise profiles $u(r=r_1, z)$ for the cases exhibiting a power-law exponent of $-3/5$ and $-1$, respectively.
The values of $C_a$ and $C_d$ are determined using the relation $u(r_1, 0) = C_a f'(0) r_1^{-3/5}$ and $u(r_1,0) = C_d/r_1$, which are subsequently used to determine the boundary layer thickness $\delta_{a,d}(r_1)$ for that profile.
When the profiles are rescaled according to \eqref{eqn:uansatz} or \eqref{eqn:u2ansatz}, and plotted against the similarity coordinate, they collapse close to a universal curves.
The theoretical profiles $f'(\xi)$ and $\text{sech}^2(\xi/\sqrt{2})$, respectively, well-approximate these universal curves.
Apart from random measurement noise, systematic departure of the data from these curves occurs due to two reasons: the return flow in the region outside the boundary layer and departures from the power-law behavior at the measurement location $r=r_1$. 
This collapse validates the thickness of the boundary layer arising from the adsorption- and dissolution-dominated regimes.

{\it Signature 3:}
The combination of radial decay as $r^{-3/5}$ and depth-wise profiles shown in Figure \ref{fig:cabl} is only possible when driven by an adsorbed layer of surfactant spreading as $2 \pi r u(r,0) c_2(r) = q_2$, or a small perturbation thereof. 
However, the agreement in Figure \ref{fig:cabl2} of the measured velocity profile with the leading order of \eqref{eqn:u2ansatz} is not conclusive proof of the flow being driven by a dissolved surfactant.
It is so because, as explained in Ref. \cite{Mandre2017b}, Squire's radial jet \cite{Squire1955} forced by a momentum source at the origin also exhibits $r^{-1}$ decay {\it and} the velocity profile \eqref{eqn:u2ansatz} to leading order. 
Only higher order corrections to the flow in the small parameter $\delta_d/r$ distinguish between Squire's radial jet and the complete solution \eqref{eqn:u2ansatz}.
The shear rate $u_z(r, z=0)$ is such a quantity; $u_z=0$ for Squire's radial jet, and $u_z = 2u/r$ from the exact solution for dissolved surfactant driven flow by Bratukhin and Maurin \cite{Bratukhin1967}.
Based on this argument, we define the third hydrodynamic signature to be $\zeta = u_z l/u$ at $z=0$, where $l=\delta_a(r)$ if the surface velocity decays as $r^{-3/5}$, and $l=r$ if it decays as $r^{-1}$.

Figure \ref{fig:signature3} shows the experimentally measured values of $\zeta$ for all the nine cases.
As expected, for cases C1 and C5 where adsorbed surfactant dominates the dynamics, $\zeta$ is scattered around the theoretically expected value $f''(0)/f'(0) \approx 0.404$.
For the cases C2 and C6, the reduction of $\zeta$ for $r\gtrsim 8r_\text{max}$ coincides with the departure of $n$ from $-3/5$.
For the remaining cases, $\zeta$ is scattered around $2$ and not around zero, implying that the flow is driven by Bratukhin and Maurin's surfactant mechanism and not a localized momentum source near the origin.
\begin{figure}
\includegraphics{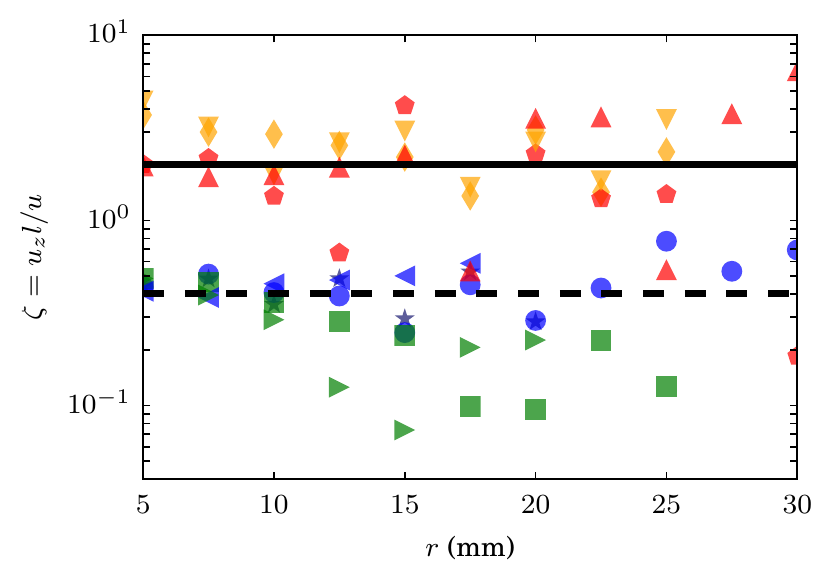}
\caption{Distribution of dimensionless shear stress on the interface. Legend same as Figure \ref{fig:expplots}.}
\label{fig:signature3}
\end{figure}

{\it Conclusion:} 
The agreement of the power-law exponent in Figure \ref{fig:expplots}, the depth-wise profile in Figures \ref{fig:cabl}-\ref{fig:cabl2}, and the dimensionless shear rate with the theoretically expected ones prove that the flow is driven by a surface stress caused by an agent transported in a manner homologous to the restrictive assumptions underlying the theoretical derivation.
Since our experimental protocol has carefully eliminated all other sources of surface stress, we are left with the unavoidable conclusion that the stress is caused by surfactant alone. 
Therefore, the surfactant dynamics within the power-law region in these cases must be as assumed in the theoretical model.
In particular, for SDS and Tergitol released on the interface at concentrations $<0.14$ CMC, the adsorbed surfactant governs the resulting dynamics, while for concentrations $>0.25$ CMC, the dissolved surfactant dynamics dominates.
A transition between the two behaviors is expected for intermediate concentrations, as suggested by the systematic deviations of $n$.
For both surfactants, the deviation of $n$ from $-3/5$ towards $-1$ at $r/r_\text{max}\gtrsim 8$ for cases C2 and C6 suggests the beginning of transition, and in the cases C3 and C7 at $r/r_\text{max} \lesssim 4$ suggests the end of the transition.
Given that the transition occurs within this range implies that the surfactant and hydrodynamic time-scales approximately overlap, rendering simple order-of-magnitude estimates unreliably to distinguish between the two regimes.
Furthermore, there is no convenient independent way to measure a pivotal parameter in characterizing the dynamics -- the fraction of the surfactant flux that is transported in an adsorbed phase. 
Therefore, using invariant hydrodynamic signatures to determine the validity of the assumptions about surfactant dynamics without {\it a priori} knowledge of the physico-chemical parameters represents a fundamental advance on the topic.

Our result is quite robust, as we demonstrated for two surfactants varying in their CMC values by factor 100, and can be used with other surfactants.
We used these signatures to determine that CA released from a gel tablet spreads in an adsorbed phase, a result that bears upon Marangoni-driven self-assembly \cite{Soh2008,Suematsu2010collective,Soh2011,Heisler2012,Kitahata2013,Grzybowski2017} and propulsion \cite{Nakata2010,Suematsu2010,Liakos2016,Akella2017}. 
Assumptions about surfactant dynamics, such as made in Ref. \cite{Roche2014}, can also be verified using the hydrodynamic signatures.
A theoretical description of the transition between the two behaviors and its dependence on the physico-chemical parameters remain to be developed.

In closing, we note a vast majority of studies \cite{Fay1969, Camp1987, Jensen1992, Jensen1995, Dussaud1998dynamics, Dussaud1998fluorescence,Fallest2010, Bandi2011} to date have focused on transient Marangoni-driven surfactant spreading dynamics, where the flow ceases once the surfactant saturates the available interface area. 
Here, we have explored the much less studied class of statistically stationary Marangoni-driven flows \cite{Bratukhin1967, Bratukhin1968, Zebib1985, Roche2014} which arise when a mechanism for surfactant outflux balances its influx rate onto the interface, thus achieving a steady-state balance. 

\acknowledgments
VSA, DKS and MMB were supported by the Collective Interactions Unit, OIST Graduate University. RSS performed the research during internship with the Collective Interactions Unit, OIST Graduate University. The authors thank Kenneth J. Meacham III for technical support, Prof. Amy Shen for help with tensiometry, Prof. Y. Yazaki-Sugiyama for help with pipette puller and Prof. Walter Goldburg for the LDV equipment.

{\it Author contributions: } SM and MMB conceived the study. MMB designed and performed the experiments with assistance from RSS, DKS and VSA, and also drafted the experimental protocol. The manuscript was drafted by SM in consultation with MMB and was approved by all authors.  




%

\clearpage
\onecolumngrid

\centerline{\bf Hydrodynamic signatures of stationary Marangoni-driven surfactant transport}
\centerline{\bf Supplemental Material}
\vspace{1mm}
\centerline{M. M. Bandi$^1$, V. S. Akella$^1$, D. K. Singh$^1$, R. S. Singh$^2$, S. Mandre$^2$}
\vspace{1mm}
\centerline{$^1$ Collective Interactions Unit, OIST Graduate University, Okinawa, Japan, 904-0495}
\centerline{$^2$ School of Engineering, Brown University, Providence, Rhode Island 02912}
~\\[2mm]

\setcounter{page}{1}
\setcounter{figure}{0}
\setcounter{equation}{0}
\renewcommand\thefigure{S\arabic{figure}} 
\renewcommand\thetable{S\arabic{table}} 

{\bf Clean room preparation: }
All experiments were performed within a static-dissipative vinyl coated (to reduce particulate matter) softwall cleanroom (5 m $\times$ 5 m) expressly converted to meet approximate class 1000 cleanroom conditions. Following thorough scrubbing of the room floor and ceiling, portable floor mount dehumidifiers and particle collectors (Terra Universal) were continuously run for 2 weeks to remove particulate matter up to 0.5 $\mu$m in size. Sticky floor mats were installed outside and inside the strip curtain entrance to the room. The room was constantly maintained at $25 \pm 1^{\circ}$C temperature.

{\bf Cleaning protocol: }
All glass components (glass syringe, petri dish, glass reservoir, and capillary) were washed in acetone followed by methanol three times and dried in an oven for 10 minutes at 100 $^{\circ}$C. They were then soaked in sulfochromic acid bath for 10 minutes, followed by a thorough rinse with de-ionized water. 
The glass components were once again baked in the oven for 30 minutes at 100 $^{\circ}$C, and irradiated in plasma to remove any residual organic impurities. PVC tubing used were washed in acetone followed by methanol three times, thoroughly rinsed with de-ionized water and dried in an oven for 20 minutes at 40 $^{\circ}$C.

{\bf Experimental preparation}:
Following initial cleaning procedures, the setup was constructed within an enclosed space modified to meet approximate class 1000 clean room conditions. 
The setup (see Figure \ref{fig:Schematic}) consisted of a square petri dish (dimensions 0.25 m $\times$ 0.25 m $\times$ 0.1 m height) constructed by gluing optically flat glass plates. 
The petri dish was filled with de-ionized (DI) water (Milli-Q resistivity 18.2 M$\Omega \cdot$cm at 25$^{\circ}$C) to 0.08 m height for measurements with SDS and Tergitol, and 0.05 m height for those with CA. 
We waited 15 minutes after filling the petri dish to allow initial transients in water current to subside. 

{\bf Capillary pulling procedure: }
Cylindrical borosilicate capillaries (World Precision Instruments) were pulled in a pipette puller (P-97, Sutter Instruments) with heating and pulling settings that resulted in a 1 cm long tapered capillary. 
This capillary was subsequently microforged to obtain a 3 -5 $\mu$m inner bore diameter, and thoroughly cleaned, and connected to the glass reservoir through PVC tubing.

{\bf Surfactant injection procedure:} 
Prior to experiments, a hole was drilled in the glass syringe piston, fit with a luer stub needle and sealed air-tight. The following procedure was followed for surfactant injection:\\
1) The glass syringe with surfactant solution was placed in a syringe pump (Harvard Apparatus PHD 22/2000). The syringe was connected to the borosilicate capillary {\it via} PVC tubing. The syringe was carefully positioned at the same height as the air-water interface to avoid surfactant flow due to hydrostatic pressure head.\\
2) The capillary was placed vertically above the air-water interface with its tapered outlet positioned 150 $\mu$m above the air-water interface.\\
3) The syringe pump was activated at a slow 0.1 $\mu$L/s constant flow rate to generate a surfactant pendant drop at the tapered capillary tip.\\
4) At the instant when the drop made contact with the interface, the surfactant was drawn by marangoni stress and the initial spreading commenced.\\
5) The air-lock was removed from the luer stub connector in the glass syringe and steady surfactant transport was allowed to be setup with an air pocket slowly forming due to surfactant withdrawal.\\
6)The syringe pump was turned off to ensure the surfactant flow was purely marangoni driven.\\
7) Once a steady surfactant flow was achieved, the capillary tip height was adjusted to bring the tapered tip in plane with the interface to minimize the surface deformation caused by the capillary bridge between the tapered tip and interface.\\
8) At the end of the experimental run, the PVC tubing was clamped and the luer stub needle was sealed once again.

Whereas the air pocket volume within syringe at the end of the experiment could provide an estimate of total surfactant flux over the experiment, it included the initial (points 6 and 7 above) as well as final (point 8 above) transient periods, which introduced large uncontrollable errors in surfactant flux estimation.

{\bf Agarose gel tablet preparation: }
Hot agarose solution (5\% weight-to-volume) in DI water was cooled between two clean glass plates, set 1 mm apart with aluminum spacers, to obtain gel sheets of uniform 1 mm thickness. Gel tablets of 3 mm diameter were punched out from the sheet (Biopunch, Ted Pella Inc.). These gel tablets were introduced in a saturated solution of CA (Wako Pure Chemical Industries, Ltd., Cat. No. 036-01002) in methanol and left for 2 hours for CA to diffuse into the gel tablets. Prior to experiments, gel tablets were rinsed in DI water to eliminate the methanol and precipitate CA in the gel matrix. 

{\bf LDV measurement details: }
We employed a legacy LDV system (TSI Inc.) with an Argon ion laser (Spectraphysics wavelength 488 nm, 35 mW continuous output) and a Bragg Cell (acousto-optic modulator) constructed in house.\\\\
\underline{Rationale for using a Bragg cell for LDV: } 
Most modern LDVs employ a beam splitter with a half-wave plate to generate two laser beams of equal intensity, which are then focused onto a Gaussian spot at a location where the velocity measurement is desired. Interference of the two beams generates a static fringe pattern at the focal point, where the fringe spacing is a function of laser wavelength. When passive colloidal tracers seeded in the flow cross the fringe pattern, they scatter light which is collected by a photodetector. The frequency of scatter pings from fringes is the doppler shifted frequency which provides a direct measure of flow velocity. Whereas at high flow velocities, this non-invasive method proves very reliable, it is prone to large velocity measurement errors when the colloidal particle either takes a long time to (low flow velocity), or does not (zero flow velocity) cross the static fringes. Since our radial velocity measurements do reach low velocities ($\sim10^{-4}$ m/s), we replace the beam splitter with a Bragg cell, which by virtue of its frequency shifting, generates traveling fringes that still provide reliable values at low to zero flow velocities.

The Bragg cell was used to generate a primary (frequency $f$) and a secondary, frequency-shifted ($f + \Delta f$, $\Delta f = 40$ MHz) beam, which were focused at a spot (90 $\mu$m diameter) causing them to interfere and generate a traveling fringe pattern. 
Colloidal tracers seeded in the flow crossed the fringe pattern and scattered light which was collected by a photodetector. 
The frequency doppler shift of the scattered light relative to fringe beating frequency provided a direct measure of flow velocity.
Polystyrene spheres of mean diameter 1.04 $\mu$m (Bangs Laboratories, Cat. No.: PS04N) were employed as colloidal passive tracers in the LDV measurement. The colloidal spheres are supplied as a suspension in water by the supplier. To ensure no impurities were transferred from the suspension to surfactants, the suspension was first washed in acetone, dried in an oven, and resuspended in de-ionized water. The new colloidal suspension was then subject to ultrasonic agitation to dissociate any colloidal clusters into individual particles. We mixed this colloidal sphere suspension (10\% solid particle fraction in water) either in the CA tablet during preparation or in de-ionized water used to prepare SDS/Tergitol stock solutions of known concentrations. 
For boundary layer measurements, we directly introduced the colloidal tracers in bulk fluid.\\\\
\underline{Radial Velocity Measurement $u(r,z=0)$:}
We used two LDV probes mounted on independent translation stages to simultaneously measure the surface radial velocity $u(r,z=0)$ in two separate sections of the range of radial distance (maximum range of $r = 0.12$ m) to be interrogated. This method was employed in order to span the full radial range in steps $\Delta r = 200~\mu$m within a duration of 10 mins, to minimize the influence of CA depletion from the tablet. The Bragg cell was employed for radial velocity measurements to assure measurement reliability as surfactant velocity fell drastically at large radial distances.
Two LDV fiber optic probes (TSI Inc., LDV 9253-120) were vertically aligned to focus their laser beams onto the air-water interface from above. The vertical position of each probe was manually adjusted with a micrometer before start of experiment to align the LDV beam focus at the interface, and the beams were aligned such that the fringes formed normal to, and traveled along the radial axis direction. The two probes mounted on two independent motorized translation stages (Newport Corp., XMS160, 160 mm travel, load capacity 100 N, on-axis accuracy: 1.5 $\mu$m) for horizontal travel along the radial direction, were simultaneously employed for measuring $u(r,z=0)$. Both motorized translation stages were reset to position at the perimeter of the tablet (for CA) or the glass capillary (for SDS and Tergitol) before start of experiment. The second probe was then moved to a distance $r = 0.0402$ m.  We waited for 2 minutes from contact of tablet or capillary at the interface for transients to die out before data collection commenced.  All data collection was automated through a LabView interface, with independent control for each probe. Each probe collected one second worth of data at a given radial position, then moved a 200 $\mu$m step along the radial direction at 300 $\mu$m/s speed, and repeated the measurement.
In this manner, the first probe scanned a radial distance $r = 0 - 0.04$ m whereas the second probe started scanning radial velocity from $r = 0.0402$ m through $r = 0.1 - 0.12$ m depending upon the experimental run. The radial velocity $u(r)$ from both probes was then patched together to reconstruct the full radial velocity profile for the spreading surfactant. No discontinuity was observed in our measurements at the $r = 0.04 - 0.0402$ m mark where measurements from both probes were patched together. Diffusivity of SDS = $1.76-4.53 \times 10^{-6}$ cm$^2$/s.

The LDV signal processor (TSI Inc., IFA 655 Digital Burst Correlator) with bandwidth of 100 KHz detected close to 70000 particle counts per second on the surface close to source ($r \approx 0$ m). 
Owing to the diverging measurement geometry, the particle counts fell with increasing radial distance, yielding about 30000 particle counts per second at the maximum radial distance ($r \simeq 0.1$ m) at which measurements were conducted. The measurement error we report is the worst case at maximum radial distance. 
Since the colloidal particles arrive at the LDV measurement spot at random, assuming Poisson statistics yields a measurement error of $1/\sqrt{N} \sim 1/\sqrt{30000} = 0.0057$ or 0.57\%.\\\\
\underline{Boundary Layer Measurement $u(r=r_1,z)$:}
A single probe sufficed for measuring the depth-wise profile $u(r = r_1, z)$ at a fixed radial position $r = r_1$ and spanning a vertical distance $z=0$ to -9 mm in steps $\Delta z = 250~\mu$m. For boundary layer profile measurements, where colloidal particles were seeded in bulk fluid as well as the surfactant, the particle counts fell from about 45000 counts per second at $z=0$, to about 7000 counts per second at maximum measurement depth of $z = - 9$ mm. Once again, assuming Poisson statistics, we obtain a measurement error of $1/\sqrt{7000} = 0.0119$ or about 1.2\%. The Bragg cell was employed for boundary layer measurements as well because the velocity magnitude falls drastically as one moves from the surface into the bulk.\\\\
\underline{Shear Stress Measurement $u_z(r,z=0)$:} 
Measurement of the shear stress requires measurement of radial velocity at two vertical locations along the $z$ axis as close to each other as the the measurement can permit. The finite cross-section of the Gaussian spot (diameter $\sim 90 \mu$m) where the two LDV beams intersect limits this distance. Preliminary tests yielded a high error for measurements made at least $90 \mu$m apart. We therefore resorted to a an unconventional method outlined below to measure shear stress:\\
1) Two LDV probes were employed, one mounted vertically above the petri dish looking down onto the interface, and the second mounted from below the dish, looking up into the dish.\\
2) Since surface radial velocities were high enough at radial positions where shear stress was measured, the Bragg cell was disconnected and replaced with the beam splitter to generate a static fringe pattern.\\
3) The pair of laser beams for each LDV probe were intentionally misaligned to reduce the intersecting spot size down to 3 static fringes with fringe spacing $\Delta s =  2.55 \mu$m.\\
4) The intentional misalignment resulted in an elliptical beam intersection spot of length $5.1 \mu$m along the radial direction and $2 \mu$m along vertical direction.\\
5) The two LDV probes were carefully positioned such that the vertical distance between the spots was $1 \mu$m, thus yielding a vertical measurement distance of $5 \mu$m.\\
6) Since the LDV probes were already designed to collect scattered light in the backscatter mode, there was no interference from forward scattered light from one LDV's spot in the other LDV's photodetector.\\
7) The LDV signal processor (TSI Inc., IFA 655 Digital Burst Correlator) was disconnected and the photodetector outputs from the LDV probes were directly fed into two independent channels of a LabView Data Acquisition System which recorded the raw time series of the photodetector outputs. In other words, instead of frequency domain analysis on doppler frequency, we performed time domain analysis on the raw output time series from the photodetectors.\\
8) The passage of a colloidal tracer registered as a scatter event with three peaks closely separated in time ($\Delta t$). The velocity was then directly obtained by dividing fringe spacing $\Delta s$ with $\Delta t$, $u = \Delta s/\Delta t$.\\
9) A total of $10^6$ such measurements were made to obtain velocity values of 0.1\% precision. The finite difference $\Delta u = u_r(r=r_1,z=0) - u_r(r=r_1,z = -5 \mu \text{m})$ divided by total vertical spacing of $5 \mu$m provided the shear stress for data presented in figure 5 of the main text.

{\bf Theoretical explanation for the scatter in $u_z$}:

This measurement of $u_z$ is especially susceptible to inaccuracies because it is O($u/r$) in magnitude and exists in the presence of a strong second derivative $u_{zz}$ which is O($u/\delta_d^2$).
The finite-difference procedure yields
$$u_z(r, z=0) \approx \dfrac{u(r,0)-u(r,-h)}{h} + \dfrac{|h| u_{zz}(r,z=0)}{2} + \dfrac{\epsilon}{|h|} + \dots,$$ 
where $h=5~\mu$ m is the finite difference spacing and $\epsilon$ represents the measurement error in the velocity. 
The error is minimized for $|h| \propto \sqrt{\epsilon/u_{zz}} \propto  \delta_d \sqrt{\epsilon/u}$, and the minimum error scales as $\Delta u_z = \sqrt{\epsilon u}/\delta_d$, where we consider the dissolution-dominated cases.
Given the weak magnitude of the shear $u/r$, the relative error $\Delta u_z/ u_z$ in the measurements scales as $\sqrt{\epsilon/u} \times r/\delta_d$.
The four digits of accuracy in velocimetry implies $\epsilon/u = O(10^{-3})$ and typically for our measurements $r/\delta \approx 10$, which implies that the minimum relative error is about $3 \times 10^{-1}$.
The optimal $|h|$ according to this analysis is about 10 $\mu$m.
The experimental $|h|$ = 5 $\mu$m was determined through experimental trials.
The above analysis rationalizes the need of an accurate velocimetry and a small finite-difference step $|h|$ for measuring $u_z$, thus explaining the need for the unconventional technique to measure $u_z$ and the increased scatter in those data.

{\bf Applying data correction to CA surface velocity: }

In general, the power-laws presented here and the underlying similarity solutions for the flow are only approximate representations of the exact solutions of the governing equations.
The comparison with the measurements for CA tablets require these corrections to be accounted.
While theoretical expressions for the corrections are difficult owing to the sensitive nature of their dependence on the system details, their general structure may be deduced readily.
The similarity solution is considered to merely be the first term in an asymptotic series solution, which in this case is
\begin{align}
 u(r,0) = \dfrac{1}{r^{3/5}} \left( a + \dfrac{b}{r} + \dfrac{c}{r^2} \dots \right), \label{eqn:approxseries}
\end{align}
where the leading order $a=K_2^{2/5} \nu^{1/5} f'(0)$ is derived from the similarity solution.
The higher order constants $b$ and $c$ depend on the specific system details that cause these corrections, but the general structure of the series in powers of $r$ is generic.
The series may be alternatively written as 
\begin{align}
 u(r,0) = \dfrac{1}{(r-\Delta r)^{3/5}} \left( a + \dfrac{c}{(r-r_0)^2} \dots \right), \label{eqn:approxseries2},
\end{align}
where the first order correction is absorbed in the leading order by simply shifting the coordinate. 
We determine the constants $a$, $\Delta r$, and $c$ by fitting \eqref{eqn:approxseries2} with the experimental data; we find that the best fits to be 
$a = 7.5\times 10^{-4}$m$^{8/5}$/s, $\Delta r = 1.50 \times 10^{-3}$ m, with a maximum difference between the experimental data and the fit of  
100 $\mu$m/s. 
The difference is too small to discern the next term in the series, i.e. $c \approx 0$ from our data.
Therefore, we plot $u(r,0)$ against $r-\Delta r$ for the flow driven by CA to account for the finite size of the source.

We note that a similar attempt to fit the measured surface velocity to the corrected version of \eqref{eqn:dispowerlaw} does not yield such a good fit.

\begin{figure}
\includegraphics{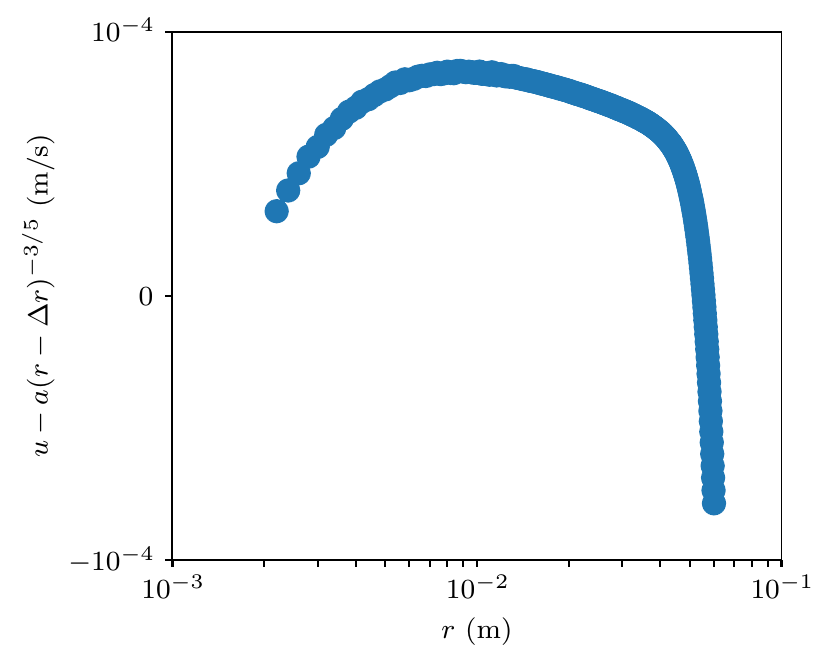}
\caption{Difference between the experimentally measured $u(r,0)$ for CA and the fit \eqref{eqn:approxseries2}. }
\label{fig:expplots2}
\end{figure}


\end{document}